\documentclass[aps,prd,twocolumn,nofootinbib,superscriptaddress,showkeys]{revtex4}
\usepackage{amssymb,amsmath,graphicx,enumerate}
\usepackage{subfigure}
\usepackage{dcolumn,booktabs}

\usepackage[dvipdfm,colorlinks=true,citecolor=blue,pdfstartview=FitH]{hyperref}

\usepackage{color,framed} 
\definecolor{shadecolor}{rgb}{0.3,0.3,1} 

\def\bea{\begin{equation}}
\def\eea{\end{equation}}
\newcommand{\thbs}{triply heavy baryons}

\newcommand{\thbf}{$\Omega_{ccc}$ and $\Omega_{bbb}$}
\newcommand{\rt}{Regge trajectory}
\newcommand{\rts}{Regge trajectories}
\newcommand{\tr}{trajectory}
\newcommand{\trs}{trajectories}

\newcommand{\gev}{{\rm GeV}}
\newcommand{\mev}{{\rm MeV}}
\newcommand{\omegabc}{$\Omega_{ccc}$ and $\Omega_{bbb}$}

\newcommand{\cltb}{$\bar{3}_c$}
\newcommand{\cltba}{\bar{3}_c}

\begin{document}
\title{$\lambda$-$\rho$ mode mixing and radial Regge trajectories for triply heavy baryons $\Omega_{ccc}$ and $\Omega_{bbb}$ in the diquark picture}
\author{Jiao-Kai Chen}
\email{chenjk@sxnu.edu.cn, chenjkphy@outlook.com (corresponding author)}
\affiliation{School of Physics and Electronic Engineering, Shanxi Normal University, Taiyuan 030031, China}
\author{Xin-Ru Liu}
\email{1170394732@qq.com}
\affiliation{School of Physics and Electronic Engineering, Shanxi Normal University, Taiyuan 030031, China}
\author{Qi Liu}
\email{18803429267@163.com}
\affiliation{School of Physics and Electronic Engineering, Shanxi Normal University, Taiyuan 030031, China}


\begin{abstract}
The mixed-mode trajectories arising from permutation-enforced $\lambda$-$\rho$ mode mixing are first investigated in the diquark picture. By accounting for the internal structure and substructure of baryons, we propose an approach to construct the mixed-mode Regge trajectories for baryons composed of three identical quarks. For triply heavy baryons $\Omega_{ccc}$ and $\Omega_{bbb}$, naive diquark-quark basis states without full permutation symmetrization violate the exchange symmetry required by the Pauli exclusion principle for three identical fermions; consequently, physical eigenstates arise as permutation-enforced $\lambda$-$\rho$ mode admixtures. Under the minimal-mixing assumption, we explicitly perform permutation symmetrization for the $J^{P}=(3/2)^{+}$ radially-excited S-wave states and construct mixed-mode Regge trajectory formulas for these symmetrized states.
Unlike triply heavy baryons $\Omega_{ccb}/\Omega_{cbb}$ in the fixed diquark $(cc)b/(bb)c$ picture where well-defined pure $\rho$- and $\lambda$-trajectories exist, genuine pure-mode trajectories no longer survive for $\Omega_{ccc}$ and $\Omega_{bbb}$ due to permutation-driven cross-mode mixing. The resulting effective radial trajectories exhibit a novel sprouting-branch-like behavior. Both the original unmixed trajectories and the mixed-mode effective trajectories exhibit behavior of $M\sim x^{2/3}$ (where $x=n_r,\,N_r,\,n_r+N_r$). Radially excited masses are crudely estimated via the mixed-mode Regge trajectories, which show reasonable agreement with existing theoretical predictions.
\end{abstract}

\maketitle


\section{Introduction}

Heavy hadrons serve as valuable laboratories for probing the nonperturbative dynamics of Quantum Chromodynamics (QCD). The triply-heavy baryons $\Omega_{ccc}$ and $\Omega_{bbb}$, composed of three identical heavy quarks ($ccc$ or $bbb$), have not yet been observed experimentally \cite{pdg:2026}. Nevertheless, they provide an ideal theoretical test-bed for investigating quark confinement, the diquark picture, and three-body hadron dynamics, and they have been studied using a vast variety of approaches \cite{Hasenfratz:1980ka,Mathur:2018epb,Faustov:2021qqf,Najjar:2025dzl,
Chen:2026wxd,Vijande:2015faa,Wei:2016jyk,Yang:2019lsg,Liu:2019vtx,Serafin:2018aih,Zhou:2025fpp,
Shah:2017jkr,Bokade:2025ldl,Brown:2014ena,Llanes-Estrada:2011gwu,Dhindsa:2024erk,
Yu:2025gdg,deArenaza:2024dhe,Jia:2006gw,Zhang:2009re,Salehi:2022lkt,FolcoCapossoli:2019imm,
Padmanath:2013zfa,Meinel:2012qz,Martynenko:2007je,Mohan:2026rcg,Yin:2021uom,Silvestre-Brac:1996myf}.

In the diquark picture, triply heavy baryons possess two distinct classes of excitation modes. The $\rho$-mode describes excitations inside the diquark subsystem, while the $\lambda$-mode corresponds to relative excitations between the whole diquark and the remaining heavy quark. For triply heavy baryons with nonidentical constituents, such as $\Omega_{ccb}$ in the configuration $b(cc)$ and $\Omega_{cbb}$ in the configuration $c(bb)$ (where $(cc)$ and $(bb)$ denote diquarks), the $\rho$-mode and $\lambda$-mode can be decoupled, and separate Regge trajectories can be constructed for each mode \cite{Xie:2024lfo}. For systems with three identical quarks, namely $\Omega_{ccc}$ and $\Omega_{bbb}$, the Pauli exclusion principle imposes strict permutation-symmetry constraints on the total wave function. The $\rho$-mode and $\lambda$-mode basis states without symmetrization do not respect the permutation symmetry for three identical particles. Consequently, physical eigenstates must be superpositions of $\rho$-mode and $\lambda$-mode components, giving rise to $\lambda$-$\rho$ mode mixing \cite{Faustov:2021qqf,
Serafin:2018aih,Bokade:2025ldl}.

Most existing theoretical studies on Regge trajectories for triply heavy baryons mainly focus on the $\lambda$-trajectory alone \cite{Shah:2017jkr,Cakir:2026fzd,Chen:2023djq,Yu:2025gdg}. Few studies address both the $\lambda$- and $\rho$-{\trs} \cite{Ishida:1994yt,Xie:2024lfo}. To our knowledge, no prior works treat mixed-mode trajectories.
In this paper, we investigate radial {\rts} for $\Omega_{ccc}$ and $\Omega_{bbb}$ in the diquark picture. Starting from the Pauli-exclusion-principle-imposed permutation-symmetry constraints, we implement explicit permutation symmetrization under the minimal-mixing assumption to handle permutation-enforced $\lambda$-$\rho$ mode admixtures for the $J^{P}=(3/2)^+$ radially-excited S-wave states. We then construct mixed-mode Regge trajectory formulas and crudely estimate masses of radial excitations. The effective mixed-mode trajectories exhibit a novel sprouting-branch-like behavior.

The paper is organized as follows. Sec. \ref{sec:rgr} present the procedures to obtain the mixed-mode Regge trajectory relations. In Sec. \ref{sec:rtdiquark}, we perform numerical calculations, analyze the novel sprouting-branch-like behavior of mixed-mode trajectories, and compare our results with other theoretical predictions. Conclusions are given in Sec. \ref{sec:conc}.

\section{{\rt} relations}\label{sec:rgr}

\subsection{Preliminary}\label{subsec:prelim}

\begin{figure}[!phtb]
\centering
\includegraphics[width=0.25\textheight]{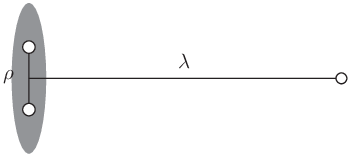}
\caption{Schematic diagram of a baryon in the diquark-quark picture. The left gray part represents a diquark composed of quark 1 and quark 2. The right circle denotes quark 3.}\label{fig:tr}
\end{figure}

In the diquark picture, baryons consist of one quark in color $3_c$ and one diquark in color $\bar{3}_c$; see Fig. \ref{fig:tr}. $\rho$ separates two quarks inside the diquark, while $\lambda$ separates the remaining quark and the diquark cluster. There exist two distinct excitation modes: the $\rho$-mode involves the radial and orbital excitation in the diquark, whereas the $\lambda$-mode involves the radial or orbital excitations between the quark and the diquark. Consequently, there exist two series of {\rts}: one series of $\rho$-{\trs} and one series of $\lambda$-{\trs}.

In the diquark picture, a baryon state is denoted as
\bea\label{tetnot}
\left((Q_1Q_2)^{\cltba}_{n^{2s_{12}+1}l_{s_D}}
{Q_3}^{3_c}\right)^{1_c}_{N^{2S+1}L_J},
\eea
where the superscripts {\cltb}, $3_c$ and $1_c$ are omitted for brevity. The notation in Eq. (\ref{tetnot}) is rewritten as $|n^{2s_{12}+1}l_{s_D},N^{2S+1}L_J\rangle$. The diquark $(Q_1Q_2)$ is $\{Q_1Q_2\}$ or $[Q_1Q_2]$. $\{Q_1Q_2\}$ and $[Q_1Q_2]$ indicate the permutation symmetric and antisymmetric flavor wave functions, respectively. $N=N_{r}+1$ with $N_{r}=0,\,1,\,\cdots$, and $n=n_{r}+1$ with $n_{r}=0,\,1,\,\cdots$. Here $N_r$ and $n_{r}$ are the radial quantum numbers for the $\lambda$-mode and the $\rho$-mode, respectively.
$\vec{J}=\vec{L}+\vec{S}$, $\vec{S}=\vec{s}_D+\vec{s}_{Q_3}$, and $\vec{s}_D=\vec{s}_{12}+\vec{l}$.
$\vec{J}$, $\vec{s}_D$ and $\vec{s}_{Q_3}$ are the spins of baryon, diquark and quark $Q_3$, respectively. $\vec{S}$ is the summed spin of diquark and quark $Q_3$ in the baryon. $L$ and $l$ are the orbital quantum numbers for the $\lambda$-mode and the $\rho$-mode, respectively. $\vec{s}_{12}$ is the summed spin of quark 1 and quark 2 inside the diquark.

\subsection{Mixing of $\lambda$ and $\rho$ modes}

The total wave function for a baryon or diquark is written as \cite{Fayyazuddin:1994wh}
\bea\label{dwf}
\Psi=\xi_{color}\otimes\eta_{flavor}\otimes\chi_{spin}\otimes\phi_{space},
\eea
where $\xi_{color}$, $\eta_{flavor}$, $\chi_{spin}$ and $\phi_{space}$ stand for the color, flavor, spin, and spatial wave functions, respectively. For a system composed of identical quarks (including the diquarks $(bb)$ and $(cc)$, as well as the baryons $\Omega_{ccc}$ and $\Omega_{bbb}$), the total wave function must be antisymmetric due to the Pauli exclusion principle. For the baryons $\Omega_{ccc}$ and $\Omega_{bbb}$, Pauli principle should be taken into account not only for the diquarks $(cc)$ and $(bb)$ but for the entire baryons.
For diquark $(Q_1Q_2)^{\cltba}$ $(Q_1=Q_2=b,\,c)$, the color wave function is antisymmetric and reads
\begin{align}
\xi_{color}=&[r,g], \,[g,b],\, [b,r],\nonumber\\
[x,y]=&\frac{1}{\sqrt{2}}\left(xy-yx\right),\;\;x,y{\in}\{r,\,g,\,b\}.
\end{align}
The flavor wave function is symmetric and reads
\bea
\eta_{flavor}=QQ \;(Q=b,c).
\eea
The spin wave function is given by:
\begin{align}
\chi_{A}=&\frac{1}{\sqrt{2}}|\uparrow\downarrow-\downarrow\uparrow\rangle \; \text{for spin singlet state},\nonumber\\
\chi_{S}^{1,0,-1}=&|\uparrow\uparrow\rangle,\;\frac{1}{\sqrt{2}}|\uparrow\downarrow+\downarrow\uparrow\rangle, \; |\downarrow\downarrow\rangle \;\text{for spin triplet states}.
\end{align}
The spatial wave function $\phi_{nlm}$ is symmetric as $l$ is even and is antisymmetric as $l$ is odd.

In diquark picture, baryon $(Q_1Q_2)Q_3$ consists of a diquark $(Q_1Q_2)$ and a heavy quark $Q_3$, where $Q_1=Q_2=Q_3=b,\;c$. The color wave function is antisymmetric, which is
\bea
\xi_{color}=\frac{1}{\sqrt{6}}\left(rgb-grb+brg-rbg+gbr-bgr\right).
\eea
The flavor wave function is symmetric and reads
\bea
\eta_{flavor}=QQQ \;(Q=b,c).
\eea
The spin wave functions for $S=3/2$ emerging from the combination of spin 1 and spin $1/2$ are fully symmetric with respect to exchange any pair of quarks. They are denoted as $\chi_S$ and are written as
\begin{align}
\chi^{3/2}_{S}=&|\uparrow\uparrow\uparrow\rangle,\nonumber\\
\chi^{1/2}_{S}=&\frac{1}{\sqrt{3}}|\uparrow\uparrow\downarrow+\uparrow\downarrow\uparrow
+\downarrow\uparrow\uparrow\rangle,\nonumber\\
\chi^{-1/2}_{S}=&\frac{1}{\sqrt{3}}|\downarrow\downarrow\uparrow+\downarrow\uparrow\downarrow
+\uparrow\downarrow\downarrow\rangle,\nonumber\\
\chi^{-3/2}_{S}=&|\downarrow\downarrow\downarrow\rangle.
\end{align}
The spin wave functions for $S=1/2$ emerging from the combination of spin 1 an spin $1/2$ have mixed symmetry, which are symmetric in quark 1 and quark 2. They are denoted as $\chi_{MS}$. The explicit form reads
\begin{align}
\chi^{1/2}_{MS}=&\frac{1}{\sqrt{6}}|-(\uparrow\downarrow+\downarrow\uparrow)\uparrow
+2\uparrow\uparrow\downarrow\rangle,\nonumber\\
\chi^{-1/2}_{MS}=&\frac{1}{\sqrt{6}}|(\uparrow\downarrow+\downarrow\uparrow)\downarrow
-2\downarrow\downarrow\uparrow\rangle.
\end{align}
Another spin-$1/2$ doublet results in the combination of spin 0 and spin $1/2$, denoted as $\chi_{MA}$. It has mixed antisymmetry and reads
\begin{align}
\chi^{1/2}_{MA}=&\frac{1}{\sqrt{2}}|(\uparrow\downarrow-\downarrow\uparrow)\uparrow\rangle,\nonumber\\
\chi^{-1/2}_{MA}=&\frac{1}{\sqrt{2}}|(\uparrow\downarrow-\downarrow\uparrow)\downarrow\rangle.
\end{align}
The spatial wave function $\phi_{nlmNLM}$ is abbreviated as $\phi_{nlNL}$. $N$, $L$, $n$, and $l$ are defined in subsection \ref{subsec:prelim}. $n_r$ and $N_r$ are radial quantum numbers for $\rho$-mode and $\lambda$-mode, respectively; $l$ and $L$ are orbital quantum numbers for $\rho$-mode and $\lambda$-mode, respectively. States $\phi_{nlNL}$ do not have definite angular momentum, which can couple to states with angular momentum $l+L,\,l+L-1,\cdots,|l-L|$.

We now proceed to construct wave functions for baryon QQQ (more details can be found in Refs. \cite{Serafin:2018aih,Faustov:2021qqf,Bokade:2025ldl}). For baryons composed of three identical quarks, most of states are mixtures of $\lambda$-excited and $\rho$-excited states. The ground state $\phi_{1s1S}$ is fully symmetric in momentum, so the spin wave function must be symmetric. Consequently, we have
\bea
|1^{3}s_{1},1^{4}S_{3/2}\rangle=\phi_{1s1S}\otimes\chi_S.
\eea
Mixing between the $\lambda$-mode and $\rho$-mode takes place for excited states. For once-excited states, the fully symmetric state is $\phi_{1p1S}\otimes\chi_{MA}-\phi_{1s1P}\otimes\chi_{MS}$, yielding baryon spin multiplets $(3/2)^-\oplus(1/2)^-$. For higher excited states, mixing takes place not only between $\lambda$-mode and $\rho$-mode but also between radially excited states and orbitally excited states. For example, wave function $\phi_{2s1S_-}\otimes\chi_{MS}-\phi_{1p1P_{L=0}}\otimes\chi_{MA}$ gives $(1/2)^+$ state, where $\phi_{2s1S_-}=(\phi_{2s1S}-\phi_{1s2S})/\sqrt{2}$. The radially excited S-wave states with $(3/2)^+$ are
\begin{eqnarray}\label{mixst}
&|n^{3}s_{1},N^{4}S_{3/2}\rangle_+=\phi_{nsNS_+}\otimes\chi_S, \nonumber\\
&\phi_{nsNS_+}=\left\{\begin{array}{cc}
\frac{1}{\sqrt{2}}\left(\phi_{nsNS}+\phi_{NsnS}\right), & n{\ne}N, \\
\phi_{nsNS}, & n=N.
\end{array}\right.
\end{eqnarray}
Eq. (\ref{mixst}) gives the minimal-mixing prescription; more mixed terms are not considered for higher excited states. For instance, $|3^{3}s_{1},1^{4}S_{3/2}\rangle_+$ can mix with $|2^{3}s_{1},2^{4}S_{3/2}\rangle$ and $|1d,1D_{3/2}\rangle$ \cite{Bokade:2025ldl}.
The spin-independent masses corresponding to states $|n^{3}s_{1},N^{4}S_{3/2}\rangle_+$ (Eq. (\ref{mixst})) read
\bea\label{mixstate}
M_{nsNS_+}=\frac{1}{2}\left(M_{nsNS}+M_{NsnS}\right),
\eea
where $M_{nsNS}$ and $M_{NsnS}$ are masses belonging to states $|n^{3}s_{1},N^{4}S_{3/2}\rangle$ and $|N^{3}s_{1},n^{4}S_{3/2}\rangle$, respectively.

\begin{table*}[!pthb]
\caption{Mass splittings $\Delta{M}=M_{J_1}-M_{J_2}$ of S-wave states ({\rm MeV}). $J_1=s_1+s_2$, $J_2=s_1-s_2$, $s_1{>}s_2$, where $J_1$ and $J_2$ are spins of meson and baryons. $s_1$ is spin of antiquark in meson or diquark in baryon, $s_2$ is spin of quark in mesons or baryons. The superscript $\dag$ denotes the lattice predictions from Ref. \cite{Mathur:2018epb}. }  \label{tab:massdif}
\centering
\begin{tabular*}{0.9\textwidth}{@{\extracolsep{\fill}}ccccccc@{}}
\hline\hline
  ${J_1}$        & ${J_2}$  & Refs. \cite{pdg:2026,Mathur:2018epb}   &  Refs. \cite{Ebert:2002pp,Faustov:2021qqf} &  Refs. \cite{Segovia:2013wma,Segovia:2016xqb,Ortega:2020uvc,Yang:2019lsg}  & Ref. \cite{Chen:2026wxd} & Ref. \cite{Silvestre-Brac:1996myf}  \\
\hline
 $J/\psi(1S)$   &$\eta_c(1S)$   & 112.81  & 117 &106 &121\\
 $\psi(2S)$      &$\eta_c(2S)$   & 48.297  & 98  &60  &46\\
 $\psi(3S)$      &$\eta_c(3S)$   &         & 97  &43 \\
 $\Upsilon(1S)$  &$\eta_b(1S)$   & 61.7    & 60  &47 &57\\
 $\Upsilon(2S)$  &$\eta_b(2S)$   & 24.4    & 30  &25 &14\\
 $\Upsilon(3S)$  &$\eta_b(3S)$   &         & 27  &19\\
 $B^{\ast}_c(1S)$     &$B_c(1S)$      & 55$^\dag$   & 62 &51 \\
 $B^{\ast}_c(2S)$     &$B_c(2S)$      &             & 46 &30   \\
 $B^{\ast}_c(3S)$     &$B_c(3S)$      &             & 42 &   \\
 $\Omega^{\ast}_{ccb}(1S)$     & $\Omega_{ccb}(1S)$   & 21$^\dag$   & 15 &19 &36 &37 \\
 $\Omega^{\ast}_{ccb}(2S)$ & $\Omega_{ccb}(2S)$  &         & 7  &13  &20         &15 \\
 $\Omega^{\ast}_{cbb}(1S)$     & $\Omega_{cbb}(1S)$   & 17$^\dag$   & 19 &21 & 41&34 \\
 $\Omega^{\ast}_{cbb}(2S)$ & $\Omega_{cbb}(2S)$ &          & 10 &15 & 27         &18 \\
\hline\hline
\end{tabular*}
\end{table*}

\subsection{Spin-dependent contributions }\label{subsec:spin}
Masses computed using Eqs. (\ref{mnost}), (\ref{combrt}), and (\ref{rtibaryon}) are spin-independent. Inspection of Table \ref{tab:massdif} shows that mass splittings $\Delta{M}$ for triply heavy baryons are substantially smaller than meson mass splittings, on the order of several tens of {\mev}, and diminish as the radial quantum number grows. For $(3/2)^+$ states, the spin-spin contribution is $\Delta{M}/3$, ranging from several tens of {\mev} down to a few {\mev}. These spin-dependent corrections are too small to alter the overall qualitative behavior of Regge trajectories; consequently, we neglect spin-dependent contributions throughout this work.

\subsection{{\rt} relations}
In the relativistic regime, {\rts} for hadrons composed of light constituents can be well described by the famous linear formula $M^2=\alpha{x}+\alpha_0$, where $x=J,\,n_r$ \cite{Chew:1961ev,Chew:1962eu,Gross:2022hyw}.
In practice, {\rt} functional forms differ across regimes \cite{Chen:2021kfw,Chen:2022flh}. In particular, {\rts} for hadrons composed of heavy constituents exhibit pronounced nonlinear behavior within the nonrelativistic regime \cite{Chen:2018hnx}.
In Refs. \cite{Xie:2024dfe,Xie:2024lfo}, using the spinless Salpeter equation together with the Cornell potential, we propose a general {\rt} ansatz
\begin{align}\label{massfinal}
M=&m_R+\beta_x(x+c_{0x})^{\nu}\,\,(x=l,\,n_r,\,L,\,N_r),\nonumber\\
m_R=&m_1+m_2+C',\quad \beta_x=c_{fx}c_xc_{c},
\end{align}
where ${\nu}$, $c_x$ and $c_{c}$ are listed in Table \ref{tab:eparam}. $c_{fx}$ theoretically equal unity but are treated as fitted parameters in practical calculations. $c_{0x}$ differ for distinct Regge trajectories. $\mu$, $C'$, and $\sigma'$ are
\begin{align}\label{rdmu}
\mu=&m_1m_2/(m_1+m_2),\nonumber\\
C'=&\left\{\begin{array}{cc}
C/2, & \text{diquarks}, \\
C, & \text{baryons},
\end{array}\right.
\;\sigma'=&\left\{\begin{array}{cc}
\sigma/2, & \text{diquarks}, \\
\sigma, & \text{baryons}.
\end{array}\right.
\end{align}
$\sigma$ is the string tension. $C$ is a fundamental parameter \cite{Gromes:1981cb,Lucha:1991vn} in the Cornell potential \cite{Eichten:1974af}.
Eq. (\ref{massfinal}) can be employed to discuss various systems including heavy-heavy systems,  heavy-light systems, and light-light systems: diquarks, mesons, baryons, triquarks, tetraquarks, pentaquarks, and hexaquarks \cite{Feng:2023txx,Liu:2026btq,Chen:2018hnx,Xie:2024lfo,Song:2024bkj,
Liu:2026vpi,Xie:2024dfe,Chen:2025fyh,Song:2025cla,Liu:2026npi}.

\begin{table}[!phtb]
\caption{Coefficients for heavy-heavy systems (HHS), heavy-light systems (HLS), and light-light systems (LLS).}  \label{tab:eparam}
\centering
\begin{tabular*}{0.47\textwidth}{@{\extracolsep{\fill}}cccc@{}}
\hline\hline
                   & HHS &  HLS & LLS  \\
\hline
$\nu$    & $2/3$ & $1/2$  & $1/2$  \\
$c_c$    & $\left({\sigma'^2}/{\mu}\right)^{1/3}$    & $\sqrt{\sigma'}$ & $\sqrt{\sigma'}$   \\
$c_{l,\,L}$    & $3/2$ & $2$ & $2\sqrt{2}$  \\
$c_{n_r,\,N_r}$ & ${\left(3\pi\right)^{2/3}}/{2}$      & $\sqrt{2\pi}$   & $2\sqrt{\pi}$ \\
\hline
\hline
\end{tabular*}
\end{table}

A triply heavy baryon consists of one doubly heavy diquark $(Q_1Q_2)$ and one heavy quark $Q_3$. Neglecting mixing of $\lambda$- and $\rho$-modes, we have the {\rt} relations for triply heavy baryons from Eq. (\ref{massfinal}) \cite{Xie:2024lfo}
\begin{align}\label{t2q}
M&=m_{R{\lambda}}+\beta_{x_{\lambda}}(x_{\lambda}+c_{0x_{\lambda}})^{2/3}\;(x_{\lambda}=L,\,N_r),\nonumber\\
M_{\rho}&=m_{R\rho}+\beta_{x_{\rho}}(x_{\rho}+c_{0x_{\rho}})^{2/3}\;(x_{\rho}=l,\,n_{r}),
\end{align}
where
\begin{align}\label{pa2qQ}
m_{R{\lambda}}&=M_{\rho}+m_{Q_3}+C,\nonumber\\
m_{R\rho}&=m_{Q_1}+m_{Q_2}+C/2,\nonumber\\
\beta_{L}&=\frac{3}{2}\left(\frac{\sigma^2}{\mu_{\lambda}}\right)^{1/3}c_{fL},\; \beta_{N_r}=\frac{(3\pi)^{2/3}}{2}\left(\frac{\sigma^2}{\mu_{\lambda}}\right)^{1/3}c_{fN_r},\nonumber\\
\beta_{l}&=\frac{3}{2}\left(\frac{\sigma^2}{4\mu_{\rho}}\right)^{1/3}c_{fl},\; \beta_{n_r}=\frac{(3\pi)^{2/3}}{2}\left(\frac{\sigma^2}{4\mu_{\rho}}\right)^{1/3}c_{fn_r},\nonumber\\
\mu_{\lambda}&=\frac{M_{\rho}m_{Q_3}}{M_{\rho}+m_{Q_3}},\;
\mu_{\rho}=\frac{m_{Q_1}m_{Q_2}}{m_{Q_1}+m_{Q_2}}.
\end{align}
In Eq. (\ref{t2q}), $M$ is the mass of the triply heavy baryon, and $M_{\rho}$ is the mass of the doubly heavy diquark. The second relation in Eq. (\ref{t2q}) serves to compute diquark masses. The first relation in Eq. (\ref{t2q}), together with Eqs. (\ref{t2q}) and (\ref{pa2qQ}), is employed to calculate the masses of {\thbs}.

From Eqs. (\ref{t2q}) and (\ref{pa2qQ}), one obtains
\bea\label{mnost}
M=M_{\rho}+m_{Q_3}+C+\beta_{x_{\lambda}}(x_{\lambda}+c_{0x_{\lambda}})^{2/3}
\eea
when the diquark is regarded as a constituent whose internal structure is disregarded. If instead the diquark is resolved as a bound state built from two heavy quarks \cite{Feng:2023txx}, we arrive at
\begin{align}\label{combrt}
M=&m_{Q_1}+m_{Q_2}+m_{Q_3}+\frac{3}{2}C\nonumber\\
&+\beta_{x_{\lambda}}(x_{\lambda}+c_{0x_{\lambda}})^{2/3}
+\beta_{x_{\rho}}(x_{\rho}+c_{0x_{\rho}})^{2/3}
\end{align}
from Eqs. (\ref{t2q}) and (\ref{pa2qQ}).
Provided $\lambda$-$\rho$ mode mixing is neglected, Eq. (\ref{combrt}) yields two separate series of {\rts} for triply heavy baryons: namely the $\lambda$-trajectories (with $x_{\rho}$ fixed) and the $\rho$-trajectories (with $x_{\lambda}$ fixed).
From Eqs. (\ref{t2q}) and (\ref{pa2qQ}), we have the diquark $(Q_1Q_2)$ {\rt} relation
\bea\label{diqrt}
M_{\rho}=m_{Q_1}+m_{Q_2}+C/2+\beta_{x_{\rho}}(x_{\rho}+c_{0x_{\rho}})^{2/3}.
\eea
Evidently, the diquark {\rts} [Eq. (\ref{diqrt})] are not the $\rho$-{\trs} for baryons [Eq. (\ref{combrt})] because not only the constants of these two series of {\rts} are not equal (see Eqs. (\ref{combrt}) and (\ref{diqrt})) but also the parameters $\beta_{x_{\lambda}}$ and $c_{0x_{\lambda}}$ are dependent on diquark mass (see Eqs. (\ref{pa2qQ}), (\ref{fitcfxl}), and (\ref{fitcfxnr})). The diquark {\rts} play a dominant role in the $\rho$-{\trs} for baryons, so the diquark {\rts} and $\rho$-{\trs} exhibit the same behavior, $M,M_{\rho}{\sim}x_{\rho}^{2/3}$ \cite{Xie:2024lfo}.

For baryons composed of identical quarks, however, $\lambda$-/$\rho$-{\tr} mixing becomes mandatory. As a result, pure unmixed $\lambda$-trajectories and $\rho$-trajectories cease to exist; only mixed-mode trajectories emerge. For the radially excited S-wave $(3/2)^+$ states of {\omegabc}, Eq. (\ref{combrt}) is symmetrized using Eq. (\ref{mixstate}) to give
\begin{align}\label{rtibaryon}
M_{n_r{N_r}_+}=&3m_{Q}+\frac{3}{2}C \nonumber\\
&+\frac{1}{2}\left[\beta_{N_r}(N_r+c_{0N_r})^{2/3}
+\beta_{n_r}(n_r+c_{0n_r})^{2/3}\right]\nonumber\\
&+\frac{1}{2}\left[\beta_{N_r}(n_r+c_{0N_r})^{2/3}
+\beta_{n_r}(N_r+c_{0n_r})^{2/3}\right].
\end{align}
$M_{nsNS_+}$ in Eq. (\ref{mixstate}) is rewritten as $M_{n_r{N_r}_+}$. Eq. (\ref{rtibaryon}) explicitly shows that physical masses after minimal $\lambda$-$\rho$ mixing are symmetric under exchange $n_r\leftrightarrow N_r$, i.e. $M_{n_r{N_r}_+}=M_{N_r{n_r}_+}$, consistent with permutation symmetry for three identical quarks.

\section{Radial {\rts} for $\Omega_{ccc}$ and $\Omega_{bbb}$}\label{sec:rtdiquark}

In this section, we employ the effective radial Regge trajectory formula given in Eq. (\ref{rtibaryon}) to investigate radial {\rts} for $\Omega_{ccc}$ and $\Omega_{bbb}$ and to crudely estimate the masses of $J^{P}=(3/2)^+$ radially excited states.

\subsection{Parameters}
Quark masses, string tension $\sigma$, and parameter $C$ are taken from Ref. \cite{Faustov:2021qqf}. Parameters for doubly heavy diquarks $(cc)$ and $(bb)$ are imported from Ref. \cite{Feng:2023txx} and listed in Table \ref{tab:parmv}. Once these parameters are fixed, diquark masses for $\rho$-mode excitations can be evaluated via Eqs. (\ref{t2q}) and (\ref{pa2qQ}).
To compute the $\lambda$-mode states,
the parameters $c_{fL}$, $c_{fN_r}$, $c_{0{L}}$, and $c_{0N_r}$ are required. Since experimental data for $\Omega_{ccc}/\Omega_{bbb}$ are absent, these parameters are obtained from empirical fitting relations \cite{Xie:2024dfe,Liu:2026npi}
\begin{eqnarray}
c_{fL}=&1.116 + 0.013\frac{\mu_{\lambda}}{m_0},\; c_{0L}=0.540- 0.141\frac{\mu_{\lambda}}{m_0}, \label{fitcfxl}\\
c_{fN_r}=&1.008 + 0.008\frac{\mu_{\lambda}}{m_0}, \;  c_{0N_r}=0.334 - 0.087\frac{\mu_{\lambda}}{m_0},\label{fitcfxnr}
\end{eqnarray}
where $m_0=1$ {\gev}. $\mu_{\lambda}$ is the reduced mass defined in Eq. (\ref{pa2qQ}). As more experimental data or more theoretical data become available, the fitted formulas will be refined.

\begin{table}[!phtb]
\caption{Parameter values taken from Refs. \cite{Feng:2023txx,Faustov:2021qqf}.}  \label{tab:parmv}
\centering
\begin{tabular*}{0.45\textwidth}{@{\extracolsep{\fill}}cc@{}}
\hline\hline
          & $m_{c}=1.55\; {\gev}$, \; $m_b=4.88\; {\gev}$, \\
          & $\sigma=0.18\; {\gev^2}$,\; $C=-0.3\; {\gev}$, \\
          & $c_{fn_{r}}=1.0$,\; $c_{fl}=1.17$,  \\
$(cc)$    & $c_{0n_{r}}(1^3s_1)=0.205$,\quad $c_{0{l}}(1^3s_1)=0.337$,\\
$(bb)$    & $c_{0n_{r}}(1^3s_1)=0.01$,\quad  $c_{0{l}}(1^3s_1)=0.001$.\\
\hline
\hline
\end{tabular*}
\end{table}

\subsection{Radial {\trs}}\label{subsec:rts}

\begin{table}[!phtb]
\caption{Predicted masses for $(3/2)^+$ states of {\thbf} (in ${\gev}$). $M_{nsNS_+}$ in Eq. (\ref{mixstate}) is rewritten as $M_{n_r{N_r}_+}$. There is $M_{n_r{N_r}_+}=M_{N_r{n_r}_+}$. Calculations employ Eqs. (\ref{pa2qQ}), (\ref{rtibaryon}), (\ref{fitcfxl}), (\ref{fitcfxnr}) together with parameters listed in Table \ref{tab:parmv}.}  \label{tab:massnN}
\centering
\begin{tabular*}{0.48\textwidth}{@{\extracolsep{\fill}}clcc@{}}
\hline\hline
 $n_r+N_r$  &  $M_{n_r{N_r}_+}$        & $(cc)c$  &  $(bb)b$  \\
\hline
   $0$      &   $M_{00_+}$     &4.65   &14.27     \\
   $1$      &   $M_{01_+}$, $M_{10_+}$      &5.11   &14.66     \\
   $2$      &   $M_{02_+}$, $M_{20_+}$     &5.44   &14.90     \\
            &   $M_{11_+}$     &5.57   &15.04     \\
   $3$      &   $M_{03_+}$, $M_{30_+}$     &5.73   &15.10     \\
            &   $M_{12_+}$, $M_{21_+}$     &5.90   &15.28     \\
   $4$      &   $M_{04_+}$, $M_{40_+}$     &5.99   &15.28     \\
            &   $M_{13_+}$, $M_{31_+}$     &6.18   &15.48     \\
            &   $M_{22_+}$     &6.23   &15.52     \\
\hline\hline
\end{tabular*}
\end{table}

\begin{figure}[!phtb]
\centering
\subfigure[]{\label{subfigure:cfa}\includegraphics[scale=0.75]{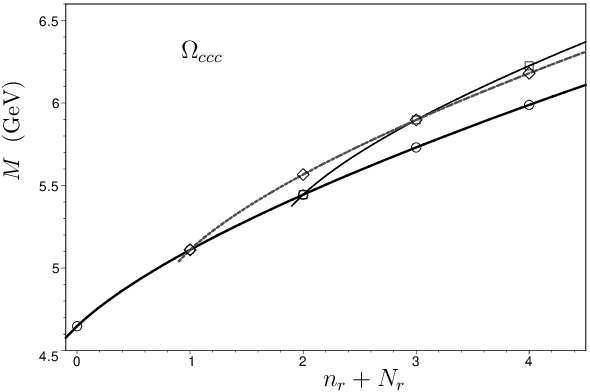}}
\subfigure[]{\label{subfigure:cfb}\includegraphics[scale=0.75]{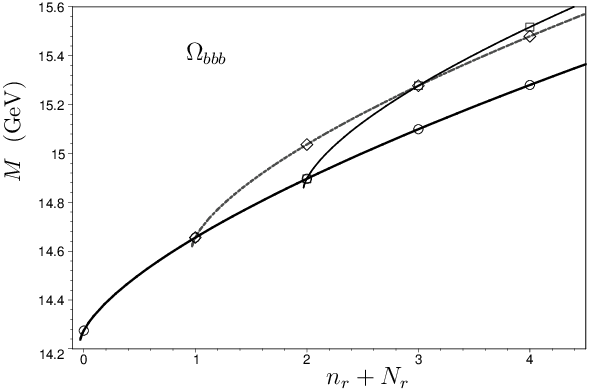}}
\caption{Permutation-symmetrized radial Regge trajectories for \(\Omega_{ccc}\) and \(\Omega_{bbb}\) after permutation-forced $\lambda$-$\rho$ mode mixing. Circles represent $M_{00_+}$, $M_{01_+}$, $M_{02_+}$, $M_{03_+}$, $\cdots$. Diamonds denotes $M_{10_+}$, $M_{11_+}$, $M_{12_+}$, $M_{13_+}$, $\cdots$. Squares represent $M_{20_+}$, $M_{21_+}$, $M_{22_+}$, $\cdots$. The thick lines in \ref{subfigure:cfa} and \ref{subfigure:cfb} correspond to Eqs. (\ref{rcca}) and (\ref{rbba}), respectively. The dashed lines in \ref{subfigure:cfa} and \ref{subfigure:cfb} correspond to Eqs. (\ref{rccb}) and (\ref{rbbb}), respectively. The thin lines in \ref{subfigure:cfa} and \ref{subfigure:cfb} correspond to Eqs. (\ref{rccc}) and (\ref{rbbc}), respectively. Data are listed in Table \ref{tab:massnN}.}\label{fig:rg}
\end{figure}

After imposing permutation symmetrization for minimal $\lambda$-$\rho$ mixing, physical $J^P=(3/2)^+$ radially excited states can be constructed according to Eq. (\ref{mixst}), with corresponding spin-independent masses supplied by Eq. (\ref{mixstate}). Spin-dependent mass corrections are neglected here, as discussed in subsection \ref{subsec:spin}.
Numerical mass predictions for these symmetrized radially excited states are calculated using the effective trajectory formula given in Eq. (\ref{rtibaryon}), together with Eqs. (\ref{pa2qQ}), (\ref{fitcfxl}), (\ref{fitcfxnr}), and parameters listed in Table \ref{tab:parmv}.
The calculated results are compiled in Table \ref{tab:massnN}, where \(M_{n_rN_r+}\) denotes masses evaluated via the symmetrized formula Eq. (\ref{rtibaryon}). By construction, the symmetry relation $M_{n_rN_r+}=M_{N_rn_r+}$ holds.

$M_{00_+}$, $M_{01_+}$, $M_{02_+}$, $M_{03_+}$, $\cdots$ lie on the {\rt} for the $\Omega_{ccc}$ states $|1^{3}s_{1},N^{4}S_{3/2}\rangle_+$. Fitting these masses gives the formula
\bea\label{rcca}
M=4.4228+ 0.5985 (0.2319+ n_r+N_r)^{2/3}, \, n_r+N_r\ge0.
\eea
In other words, the {\rt} results for $|1^{3}s_{1},N^{4}S_{3/2}\rangle_+$ states obtained using Eq. (\ref{rtibaryon}) can be well reproduced by the simple formula of Eq. (\ref{rcca}). Likewise, the points $M_{10_+}$, $M_{11_+}$, $M_{12_+}$, $M_{13_+}$, $\cdots$ populate on the {\rt} for $\Omega_{ccc}$ states $|2^{3}s_{1},N^{4}S_{3/2}\rangle_+$, with fitted formula
\bea\label{rccb}
M=4.886+ 0.5920 (-0.7661 + n_r+N_r)^{2/3}, \, n_r+N_r\ge1.
\eea
$M_{20_+}$, $M_{21_+}$, $M_{22_+}$, $\cdots$ lie on the {\rt} for $\Omega_{ccc}$ $|3^{3}s_{1},N^{4}S_{3/2}\rangle_+$, with the fitted formula
\bea\label{rccc}
M=5.2211+ 0.5875 (-1.7641 +  n_r+N_r)^{2/3}, \, n_r+N_r\ge2.
\eea
Following the same procedure for $\Omega_{ccc}$, we obtain fitted {\rt} for $\Omega_{bbb}$  $|1^{3}s_{1},N^{4}S_{3/2}\rangle_+$:
\bea\label{rbba}
M=14.2338+ 0.4131 (0.0319+ n_r+N_r)^{2/3}, \, n_r+N_r\ge0.
\eea
The fitted {\rt} for $\Omega_{bbb}$  $|2^{3}s_{1},N^{4}S_{3/2}\rangle_+$ is
\bea\label{rbbb}
M=14.616+ 0.4124 (-0.9701 + n_r+N_r)^{2/3}, \, n_r+N_r\ge1.
\eea
The fitted {\rt} for $\Omega_{bbb}$  $|3^{3}s_{1},N^{4}S_{3/2}\rangle_+$ reads
\bea\label{rbbc}
M=14.8572+ 0.4117 (-1.9701 +  n_r+N_r)^{2/3}, \, n_r+N_r\ge2.
\eea
Mass values and the above trajectories are visualised in Fig. \ref{fig:rg}. Circles mark points belonging to $|1^3s_1,N\,^4S_{3/2}\rangle_+$; diamonds correspond to $|2^3s_1,N\,^4S_{3/2}\rangle_+$; squares represent $|3^3s_1,N\,^4S_{3/2}\rangle_+$. Thick, dashed and thin solid lines denote fitted Regge trajectories for each series of states respectively.

For {\thbs} $\Omega_{bcc}$ in the configuration $b(cc)$ and $\Omega_{cbb}$ in the configuration $c(bb)$, pure $\rho$-{\trs} and $\lambda$-{\trs} can be constructed separately \cite{Xie:2024lfo}; however, for $\Omega_{ccc}$ and $\Omega_{bbb}$ composed of three identical quarks, unsymmetrized quark-diquark basis states violate exchange symmetry for three identical quarks. Consequently, physical energy eigenstates must be combinations of $\rho$-mode and $\lambda$-mode basis components, which gives rise to unavoidable $\lambda$-$\rho$ mode mixing. Owing to the permutation-induced $\lambda$-$\rho$ mixing, pure $\rho$-excited states and pure $\lambda$-excited states cease to exist for $\Omega_{ccc}$ and $\Omega_{bbb}$. As a direct consequence, separate unmixed $\rho$-{\trs} and $\lambda$-{\trs} are no longer well-defined; instead we obtain mixed-mode effective {\rts}, as embodied by Eqs. (\ref{rtibaryon}), (\ref{rcca}), (\ref{rccb}), (\ref{rccc}), (\ref{rbba}), (\ref{rbbb}), and (\ref{rbbc}).

As illustrated in Fig. \ref{fig:rg}, the mixed-mode effective Regge trajectories for $\Omega_{ccc}$ and $\Omega_{bbb}$ exhibit sprouting-branch-like behavior. Only the lowest lying family $|1^{3}s_1,N^{4}S_{3/2}\rangle_{+}$ originates at the origin $n_r+N_r=0$. Higher excited families $|2^{3}s_1,N^{4}S_{3/2}\rangle_{+}$ and $|3^{3}s_1,N^{4}S_{3/2}\rangle_{+}$ sprout successively at finite thresholds $n_r+N_r=1,\,2,\dots$, instead of starting from the common origin. Once a new family appears, its trajectory gradually diverges away from the previously existing curves with increasing total radial excitation quantum number $n_r+N_r$. This sprouting-and-diverging feature originates from Pauli-enforced $\lambda$-$\rho$ mode mixing for three identical heavy quarks together with our minimal-mixing assumption. It should be emphasized that these effective trajectories cannot be identified with the separate pure $\lambda$ and pure $\rho$ Regge trajectories for $\Omega_{ccb}/\Omega_{cbb}$ systems.

In the absence of mixing between $\lambda$-mode and $\rho$-mode, both $\rho$-trajectories and $\lambda$-trajectories for {\thbs} have the same behavior $M\sim x^{2/3}$ (Eq. (\ref{combrt})), where $x$ stands for radial/orbital quantum number. This {\rt} behavior is not immediately transparent when inspecting the symmetrized {\rt} relation given in Eq. (\ref{rtibaryon}).
Nevertheless, the simple fitted parametrisations (Eqs. (\ref{rcca}), (\ref{rccb}), (\ref{rccc}), (\ref{rbba}), (\ref{rbbb}), and (\ref{rbbc})), which approximate Eq. (\ref{rtibaryon}) for physical state series, inherit the same $M\sim x^{2/3}$ scaling behavior. Hence the effective trajectories (Eq. (\ref{rtibaryon})) for symmetrized states also display the $M\sim x^{2/3}$ behavior.
Moreover, the good approximation of Eq. (\ref{rtibaryon}) by Eqs. (\ref{rcca}), (\ref{rccb}), (\ref{rccc}), (\ref{rbba}), (\ref{rbbb}), and (\ref{rbbc})) shows that the mixed-mode {\trs} can be fitted using a simple functional form $M=m_R+\beta_x (x+c_{0})^{2/3}$ although Eq. (\ref{rtibaryon}) together with (\ref{pa2qQ}), (\ref{fitcfxl}), and (\ref{fitcfxnr}) takes a lengthy and cumbersome form.

The crude mass predictions obtained from the effective Regge trajectory approach are compared with other theoretical results in Table \ref{tab:masscomp}. Overall, our mass estimates show reasonable consistency with existing model calculations for triply heavy baryons.

\begin{table*}[!phtb]
\caption{Comparison of theoretical predictions for masses of {\thbf} (in GeV).}  \label{tab:masscomp}
\centering
\begin{tabular*}{1.0\textwidth}{@{\extracolsep{\fill}}cccccccccccc@{}}
\hline\hline
  Baryon  &  $|n^{3}s_{1},N^{4}S_{3/2}\rangle_+$
          &   This work  & Ref. \cite{Faustov:2021qqf} & Ref. \cite{Chen:2026wxd} & Ref. \cite{Najjar:2025dzl} & Ref. \cite{Yang:2019lsg} & Ref. \cite{Liu:2019vtx} & Ref. \cite{Serafin:2018aih} & Ref. \cite{Shah:2017jkr} &Ref. \cite{Bokade:2025ldl} & Ref. \cite{Brown:2014ena} \\
\hline
$\Omega_{ccc}$
   &$|1^{3}s_{1},1^{4}S_{3/2}\rangle_+$ &4.65   &4.712 &4.833   &5.06  &4.798 &4.828  &4.797  &4.806 &4.660 &4.796\\
   &$|1^{3}s_{1},2^{4}S_{3/2}\rangle_+$ &5.11   &5.137 &5.294   &5.26  &5.286 &5.285  &5.309  &5.300 &5.104 \\
   &$|1^{3}s_{1},3^{4}S_{3/2}\rangle_+$ &5.44   &      &   &      &      &  &  & &5.564 \\
   &$|2^{3}s_{1},2^{4}S_{3/2}\rangle_+$ &5.57   &5.541 &   &      &      &  &  & &5.564\\
$\Omega_{bbb}$
   &$|1^{3}s_{1},1^{4}S_{3/2}\rangle_+$ &14.27   &14.468 &14.377 &13.97 &14.396 &14.432  &14.347  &14.496 &14.200 &14.366\\
   &$|1^{3}s_{1},2^{4}S_{3/2}\rangle_+$ &14.66   &14.815 &14.804  &14.20 &14.805 &14.848  &14.832  &15.163 &14.635\\
   &$|1^{3}s_{1},3^{4}S_{3/2}\rangle_+$ &14.90   &      &  &      &       &  &  &&15.086 \\
   &$|2^{3}s_{1},2^{4}S_{3/2}\rangle_+$ &15.04   &15.123 &  &      &       &  &  &&15.086 \\
\hline\hline
\end{tabular*}
\end{table*}

\subsection{Discussions}

For \(\Omega_{ccc}/\Omega_{bbb}\), Pauli permutation symmetry forces \(\lambda\)-mode and \(\rho\)-mode basis components to admix, even in the absence of any dynamical off-diagonal Hamiltonian coupling. This behavior is qualitatively distinct from non-identical-quark triply heavy baryons such as \(\Omega_{bcc}\) (\(b(cc)\) configuration), where there is no permutation-driven cross-mode mixing. For \(\Omega_{bcc}\) (\(b(cc)\) configuration), pure $\lambda$-mode and pure $\rho$-mode states can be eigenstates. Mixing may only occur as dynamical mixing, induced by non-zero off-diagonal Hamiltonian matrix elements.

Our minimal-mixing symmetrised basis only includes pairs related by exchange $n_r\leftrightarrow N_r$. Real three-quark dynamics can induce further mixing with basis states having different total radial-quantum-number sums or different orbital angular-momentum quantum numbers; such contributions are neglected here. Consequently our effective trajectories should be regarded as a approximate phenomenological description for low-lying $J^P=(3/2)^+$ radial excitations; systematic deviations are expected for highly excited states.

From discussions in sections \ref{sec:rgr} and \ref{sec:rtdiquark}, we can conclude that accounting for the internal structure and substructure of baryons is indispensable for
constructing the mixed-mode trajectories as well as the unmixed $\rho$-trajectories; without such structural considerations, functional form and trajectory parameters can only be obtained via pure fitting against theoretical or experimental data. The Regge trajectories of internal constituents can provide guidance for the distinct series of Regge trajectories for baryons \cite{Xie:2024lfo}.

\section{Conclusions}\label{sec:conc}
The mixed-mode trajectories arising from permutation-enforced $\lambda$-$\rho$ mode mixing are first investigated in the diquark picture. 
For baryons $\Omega_{ccc}$ and $\Omega_{bbb}$ composed of three identical heavy quarks, the Pauli exclusion principle imposes strict permutation-symmetry constraints on the total many-body wave function. Diquark-quark basis states built without full permutation symmetrization violate the exchange symmetry for three identical fermions. Consequently, physical eigenstates must be superpositions of pure $\rho$-mode and $\lambda$-mode basis components, giving rise to permutation-enforced $\lambda$-$\rho$ mode mixing.

In this work, we have performed explicit permutation symmetrization to implement the minimal-mixing assumption for $J^P=(3/2)^+$ radially excited S-wave states of $\Omega_{ccc}$ and $\Omega_{bbb}$. Following this symmetrization prescription, physical spatial wave-function components are given by $\phi_{nsNS_{+}}=\big(\phi_{nsNS}+\phi_{NsnS}\big)/{\sqrt{2}}$, and the corresponding spin-independent mass is given by $M_{nsNS_{+}}=\big(M_{nsNS}+M_{NsnS}\big)/2$. This minimal-mixing approximation discards further mixing channels with higher-lying radially or orbitally excited multiplets; therefore, the obtained effective states are physical approximations.

For $\Omega_{ccb}/\Omega_{cbb}$ in the fixed diquark $(cc)b/(bb)c$ picture, no permutation-driven mixing occurs between $\rho$- and $\lambda$-modes; pure $\rho$- and pure $\lambda$-trajectories are well defined. By contrast, for $\Omega_{ccc}$ and $\Omega_{bbb}$, there exist no pure $\rho$-{\trs} or pure $\lambda$-{\trs}; only $\lambda$-$\rho$ mixed {\trs} can be defined.
We construct mixed-mode {\rt} formula for symmetrized physical states after $\lambda$-$\rho$ mixing. Both the original unmixed $\rho$-trajectories and $\lambda$-trajectories, as well as the mixed-mode trajectories for $\Omega_{ccc}$ and $\Omega_{bbb}$ have the same behavior $M\sim x^{2/3}$ (where $x=n_r,\,N_r,\,n_r+N_r$) inherited from the underlying two-body heavy-heavy confinement dynamics.
The resulting effective trajectories exhibit a sprouting-branch-like morphology. This completely new feature originates directly from permutation-induced mixing.

We propose an approach to construct the mixed-mode {\rts} for baryons composed of three identical quarks in the diquark picture. Accounting for the internal structure and substructure of baryons is indispensable for constructing the mixed-mode trajectories as well as the unmixed $\rho$-trajectories; without such structural considerations, functional form and trajectory parameters can only be obtained via pure fitting against theoretical or experimental data. The Regge trajectories of internal constituents can provide guidance to the distinct series of Regge trajectories for baryons.

\end{document}